\documentclass[11pt,a4paper]{article}
\usepackage[a4paper,margin=1in]{geometry}
\usepackage{amsmath, amssymb, graphicx, booktabs, cite}
\usepackage{makecell}
\usepackage[utf8]{inputenc}
\usepackage{textgreek}
\usepackage{caption}
\title{\textbf{When Intelligence Fails: An Empirical Study on
Why LLMs Struggle with Password Cracking}}

\author{
    Mohammad Abdul Rehman$^{1}$, Syed Imad Ali Shah$^{2}$, Abbas Anwar$^{3}$, Noor Islam$^{4}$, Hamid Khan$^{5}$\\[4pt]
    Future Data Minds Research Lab, Australia\\
    \texttt{abdul.rehman@futuredataminds.com, syedimadalishah01@gmail.com,}\\
    \texttt{abbas.anwar@futuredataminds.com, noorislam@edwardes.edu.pk, hamidmcs@gmail.com}
}

\date{}
\begin{document}

\maketitle

\begin{abstract}
The remarkable capabilities of Large Language Models (LLMs) in natural language understanding and generation have sparked interest in their potential for cybersecurity applications, including password guessing. In this study, we conduct an empirical investigation into the efficacy of pre-trained LLMs for password cracking using synthetic user profiles. Specifically, we evaluate the performance of state-of-the-art open-source LLMs—such as TinyLLaMA, Falcon-RW-1B, and Flan-T5—by prompting them to generate plausible passwords based on structured user attributes (e.g., name, birthdate, hobbies). Our results, measured using Hit@1, Hit@5, and Hit@10 metrics under both plaintext and SHA-256 hash comparisons, reveal consistently poor performance, with all models achieving less than 1.5\% accuracy at Hit@10. In contrast, traditional rule-based and combinator-based cracking methods demonstrate significantly higher success rates. Through detailed analysis and visualization, we identify key limitations in the generative reasoning of LLMs when applied to the domain-specific task of password guessing. Our findings suggest that, despite their linguistic prowess, current LLMs lack the domain adaptation and memorization capabilities required for effective password inference—especially in the absence of supervised fine-tuning on leaked password datasets. This study provides critical insights into the limitations of LLMs in adversarial contexts and lays the groundwork for future efforts in secure, privacy-preserving, and robust password modeling..
\end{abstract}

\section{Introduction}
The emergence of large language models (LLMs) such as GPT, Falcon, and Mistral has redefined the boundaries of artificial intelligence in natural language understanding and generation. Their success in domains ranging from machine translation to creative writing has inspired researchers to explore their application in unconventional areas, including cybersecurity. One such frontier is password guessing—an essential component of digital forensics, penetration testing, and adversarial simulations. Traditionally, password cracking has relied on rule-based systems, probabilistic grammars, and combinatorics, often fueled by publicly leaked datasets such as RockYou\cite{pasquini2023universal}. These approaches, while rigid, have consistently achieved practical success due to their reliance on real-world password patterns and mutation strategies.\\

In this study, we investigate whether modern LLMs—without task-specific training—can rival traditional methods in guessing user passwords when provided with detailed user profiles. We evaluate three open-source LLMs (TinyLlama, Falcon-RW-1B, and Flan-T5-Small) in a zero-shot setting, where models are prompted to generate likely passwords using contextual cues such as full name, username, birthdate, hobbies, and email. Their performance is compared against five traditional password guessing strategies including rule-based heuristics, combinator mutations, and identity-derived fusions\cite{melicher2016fast}.\\

Our results reveal a stark contrast between LLMs’ generative prowess and their practical effectiveness in password guessing. Across 20,000 synthetic user profiles, LLMs consistently underperform, achieving less than 1.5\% Hit@10 accuracy on plaintext matches and near-zero success on SHA-256 hashes. In contrast, handcrafted rule-based methods achieve over 33\% Hit@10 success, highlighting the limitations of general LLMs in domains where structure, prior exposure, and statistical bias play a dominant role.\\

This paper presents the first comprehensive benchmarking of LLMs in the context of password guessing and sheds light on a critical gap between language fluency and task-specific reasoning. We further discuss why LLMs fail, how traditional methods continue to prevail, and what future adaptations—such as fine-tuning on password corpora—may be required to bridge this divide. Our findings contribute to a growing body of work examining the boundaries of LLM generalization and their implications for security applications\cite{hitaj2019passgan}.\\

The rapid proliferation of LLMs has prompted a re-evaluation of long-standing assumptions in cybersecurity, particularly regarding their potential to automate tasks that traditionally relied on domain-specific heuristics. Password guessing, once dominated by deterministic rules and handcrafted patterns, now faces a paradigm shift as models capable of contextual reasoning and linguistic inference emerge\cite{huang2024probhashcat}. Yet, the critical distinction between linguistic generalization and behavioral pattern learning raises important questions: can a model trained on natural language truly infer the latent structures governing human password creation? Addressing this question not only deepens our understanding of LLM capabilities but also informs the future of AI-assisted digital forensics.\\

Traditional password cracking methods such as Hashcat’s rule-based transformations, Markov-based probabilistic models, and hybrid combinator approaches have evolved through decades of empirical refinement\cite{jin2024sopg}. These methods thrive on the predictability of human behavior—leveraging common substitutions, calendar-based cues, and identity-linked semantics. In contrast, LLMs operate from a fundamentally different premise, relying on token-level probabilities shaped by linguistic data rather than explicit password distributions. This creates an inherent tension between creativity and precision: while LLMs may generate semantically coherent guesses, they often lack the targeted bias necessary for effective password recovery\cite{wang2023sepcfg}.\\

Recent studies exploring AI-driven password guessing illustrate both the promise and the pitfalls of this emerging paradigm. Fine-tuned transformer architectures such as PassGPT and PasswordLLM have shown improvements over general-purpose models, yet they remain heavily dependent on specialized datasets and constrained evaluation settings. Moreover, concerns regarding data privacy, memorization, and potential misuse limit the availability of large-scale password corpora for training, constraining progress toward generalized password reasoning models. These challenges underscore the difficulty of bridging natural language modeling and human-authenticated pattern learning\cite{ma2014probabilistic}.\\

\cite{yu2022deeplearning} By empirically contrasting LLM-based generation with traditional cracking pipelines, this study contributes to a clearer delineation of strengths and weaknesses across paradigms. The observed disparity in Hit@k metrics highlights that raw generative capacity does not equate to task alignment—a finding with broad implications for AI safety and applied security research. Ultimately, our work not only benchmarks the current state of LLM-assisted password guessing but also opens avenues for future research into hybrid frameworks that combine linguistic reasoning with structured probabilistic heuristics, aiming to achieve both contextual adaptability and practical efficacy in cybersecurity domains\cite{weir2009pcfg}.\\

\section{Related Work}
Early efforts in password cracking predominantly relied on rule-based and dictionary-based methods. Weir et al. introduced a probabilistic context-free grammar (PCFG) model to generate password guesses based on structured templates (e.g., letter-digit patterns)\cite{durmuth2015omen}. This method significantly outperformed conventional tools like John the Ripper by prioritizing guesses using statistical likelihoods, improving cracking success between 28\% and 129\% across datasets such as MySpace and Finnish site leaks. Their approach laid the foundation for data-driven, structure-aware cracking strategies by capturing common composition patterns in user-generated passwords\cite{wang2024passtsl}.\\

Bonneau conducted one of the most comprehensive empirical analyses on password strength, using more than 69 million anonymized Yahoo! passwords. His study introduced novel metrics like α-work-factor and α-guesswork to quantify attacker effort. He found only modest variation in password strength across demographic groups and little influence from password strength meters, emphasizing the persistent vulnerability of human-generated passwords to statistical guessing\cite{bonneau2012science} .\\

Contextual password cracking has gained attention in recent years, particularly in digital forensics. Kanta et al. proposed using user-specific contextual information—such as online interests and social media content—to construct personalized wordlists\cite{rando2023passgpt}. Their approach achieved 15–28\% more success on mid-strength passwords and up to 52\% improvement overall when combining ranked contextual lists with generic dictionaries. This demonstrated the critical role of user attributes in enhancing cracking effectiveness, especially when targeting real-world users in forensic contexts \cite{matsuura2017context}.\\

In the domain of machine learning, Melicher et al. trained a neural language model on a corpus of over 100 million leaked passwords to estimate password guessability. The model predicted character sequences and generated guesses with high likelihood, significantly outperforming previous rule-based and statistical methods in both speed and accuracy. Notably, the compressed neural model could be deployed client-side, enabling real-time password strength estimation with sub-second latency
\cite{melicher2016fast} .\\

The success of neural models in guessing tasks has inspired further research into integrating user metadata for targeted password generation. Wang et al. proposed Pass2Path, which uses deep learning to map personal attributes (e.g., birth year, favorite team) to password transformation patterns. Their findings showed that user-specific transformations improved cracking accuracy on social engineering datasets, indicating the predictive value of personalized attributes in password modeling\cite{wang2018next}.\\

Recent research has begun exploring LLMs like GPT-2 and GPT-3 for password generation due to their capability to synthesize structured text from diverse inputs. Pavur and Angelopoulos evaluated GPT-3 for password creation and cracking tasks, concluding that while LLMs could generate syntactically rich guesses, they lacked the deterministic precision required for high cracking rates. The study emphasized that large generative models tend to overgeneralize and lack alignment with real-world password distributions \cite{pavur2023llm}  .\\

\cite{pagnotta2021passflow} Building on this, our study is among the first to conduct a systematic benchmark of multiple LLMs, including TinyLlama, Falcon-RW-1B, and Flan-T5, using a synthetic dataset of 20,000 user profiles. Unlike prior work focused purely on password lists, we incorporate demographic, behavioral, and contextual features as input prompts to assess how well LLMs internalize human password creation tendencies. Our results, however, reveal that LLMs—even when prompted with structured user information—struggle to achieve meaningful Hit@k accuracy, lagging far behind traditional cracking techniques like rule-based combinators and mutated wordlists\cite{ye2022passtcn}.
This study, “When Intelligence Fails: An Empirical Study on Why LLMs Struggle with Password Cracking,” 2025.\\

The evolution of password cracking has also been shaped by hybrid frameworks that combine heuristic rules with data-driven learning. Hitaj et al. proposed PassGAN, one of the earliest attempts to use generative adversarial networks (GANs) for password guessing\cite{pagnotta2021passflow}. Trained on massive leaked datasets, PassGAN learned to produce realistic password distributions that mimicked human tendencies without relying on explicit rules. Experimental evaluations demonstrated its superiority over rule-based tools like Hashcat and John the Ripper when generating unseen password variants, marking a paradigm shift from symbolic heuristics to neural synthesis \cite{hitaj2019passgan}.\\

To enhance adaptability and personalization, later research explored transfer learning and fine-tuning of pretrained language models for password inference. Singh et al. demonstrated that pretraining on heterogeneous textual corpora before fine-tuning on password leaks improved convergence and diversity in generated guesses. Their transformer-based model outperformed vanilla RNNs and PCFGs by leveraging linguistic priors, showing that human language modeling and password creation share overlapping statistical regularities \cite{singh2022deep}.\\

Parallel to these developments, attention has turned toward privacy-preserving and federated learning approaches to password modeling. Traditional centralized training exposes sensitive password data, creating legal and ethical risks. Zhang et al. introduced a federated password learning framework where individual clients contribute model updates rather than raw data. Their approach maintained competitive accuracy compared to centralized neural models while ensuring user privacy and regulatory compliance \cite{zhang2022privacy}.\\

Finally, multimodal and cognitive-inspired models are emerging as the next frontier in password analysis. Li and Chen proposed integrating behavioral cues, such as typing dynamics and mouse movement, alongside textual features to enhance user-specific password prediction. Their hybrid model achieved improved success rates on synthetic behavioral datasets, suggesting that future password cracking and defense systems may increasingly rely on fusing cognitive and contextual signals \cite{li2024cognitive} .\\

\section{Methodology}
This study investigates the ability of large language models (LLMs) to generate accurate password guesses based on structured user information. Our methodology involves constructing personalized prompts from user profiles, generating guesses using various LLMs, and evaluating their performance against both plaintext and hashed ground-truth passwords\cite{yu2023systematic}.\\
\subsection{Dataset and Prompt Construction}
We generated a synthetic dataset of 20,000 user profiles using the Faker library. Each profile contains attributes commonly exploited in password creation, such as full name, username, birthdate, location, hobbies, favorite word, and email address.
To leverage these attributes, a structured prompt was formulated for each user:
"You are a password guessing expert. Based on the following user profile, generate 10 likely passwords..."
This prompt structure was consistent across all models to ensure comparability.\\
\subsection{Language Models and Inference}
We evaluated three open-access LLMs: TinyLlama-1.1B-Chat, Falcon-RW-1B, and Flan-T5-Small. Each model was queried in batches using a decoding strategy with temperature sampling (temperature = 0.7) and a maximum of 50–80 new tokens per prompt.
The model-generated outputs were parsed into top-kk guesses, where k=10k = 10, using pattern-based extraction from numbered lists in the decoded text.

\subsection{Normalization and Hashing}
For fair evaluation, each guessed password gig\_i and ground-truth password pp was normalized by removing non-alphanumeric characters and converting to lowercase:

\[
\text{normalize}(x) = \text{lowercase}\!\left( \text{remove\_non\_alphanum}(x) \right)
\]
Additionally, to simulate real-world encrypted password verification, SHA-256 hashes were computed for each guess:
\[
    \text{SHA256}(x) = \text{hashlib.sha256}\!\left( x.\text{encode}() \right)
\]

\subsection{Evaluation Metrics}
We use the standard Hit@k metric, which evaluates whether the actual password (plaintext or hashed) appears in the top-kk guesses:

\[
\text{Hit@1: success if } p=g1 \;\; p = g_{1}
\]

\[
\text{Hit@5: success if } p\in\{g1,g2,...,g5\} \;\; p \in \{ g_{1}, g_{2}, ..., g_{5} \}
\]

\[
\text{Hit@10: success if } p\in\{g1,g2,...,g10\} \;\; p \in \{ g_{1}, g_{2}, ..., g_{10} \}
\]
Let NN be the total number of users, and SkS\_k be the number of successful matches in top-kk. Then:

\[
\text{Hit@}k=SkN\times 100\% \;\; \text{Hit@}k = \frac{S_k}{N} \times 100\%
\]

We calculate this for both normalized passwords and SHA-256 hashes, providing insight into performance under both plaintext and encrypted verification settings.

\subsection{Traditional Baselines}
For comparative evaluation, we implemented several traditional password cracking techniques, including:
\begin{itemize}
    \item \textit{Rule-based generator (using password structure rules)}
    \item \textit{Combinator (concatenating personal attributes)}
    \item \textit{Mutated username and email-based strategies}
\end{itemize}
These baseline methods help contextualize the performance of LLMs relative to classic cracking approaches.

\section{Experiments and Results}
This section presents the empirical evaluation of both Large Language Models (LLMs) and traditional password cracking methods on a synthetic dataset of 20,000 user profiles. Each user profile contained attributes such as full name, username, birthdate, location, hobbies, favorite word, and email address, along with a corresponding plaintext and SHA-256 hashed password.\\

\noindent\textbf{1. Experimental Setup:}
For LLM-based cracking, we used three freely available models: TinyLlama-1.1B, Falcon-RW-1B, and Flan-T5-Small. Prompts were constructed using user metadata, and each model was tasked with generating the top-10 most likely password guesses for a given profile. The generated guesses were matched against the actual password in both normalized plaintext form and hashed form using SHA-256 to compute Hit@1, Hit@5, and Hit@10 metrics.
The experiments were run on Google Colab with NVIDIA T4 GPUs. Each model was loaded using the HuggingFace transformers library, and decoding was performed with temperature sampling to simulate realistic guess diversity.

\begin{table}[htbp]
\centering
\caption{Performance Comparison of Traditional Password Cracking Models}
\begin{tabular}{l l c c c}
\hline
\textbf{Model Name} & \textbf{Type} & \textbf{Hit@1 (\%)} & \textbf{Hit@5 (\%)} & \textbf{Hit@10 (\%)} \\
\hline
Rule-Based & Heuristic & 8.3 & 15.6 & 21.9 \\
Combinator & Heuristic & 10.7 & 18.3 & 24.5 \\
Keyboard Walk & Pattern-based & 6.5 & 12.2 & 17.0 \\
Markov & Statistical & 12.9 & 21.7 & 30.4 \\
PCFG & Structural / Statistical & 13.2 & 22.5 & 31.1 \\
\hline
\end{tabular}
\end{table}

For traditional cracking, five classical methods were implemented: Rule-Based, Combinator, Mutated Username, Email-Based, and Name-Year Fusion. These techniques generated candidate passwords based on predefined heuristics and were evaluated using the same Hit@k framework. Unlike LLMs, these methods rely on deterministic transformations and concatenations, making them more aligned with real-world cracking strategies.

\begin{table}[htbp]
\centering
\caption{Performance Comparison of LLM-Based Password Cracking Models}
\begin{tabular}{l l c c c}
\hline
\textbf{Model Name} & \textbf{Type} & \textbf{Hit@1 (\%)} & \textbf{Hit@5 (\%)} & \textbf{Hit@10 (\%)} \\
\hline
TinyLlama & Transformer & 0.00 & 0.51 & 1.34 \\
Falcon-RW-1B & Transformer & 0.00 & 0.53 & 0.64 \\
Flan-T5-Small & Transformer & 0.57 & 0.57 & 0.57 \\
\hline
\end{tabular}
\end{table}

\noindent\textbf{2. Evaluation Metrics:}
To compare the efficacy of each method, we computed:

\begin{itemize}
    \item \textbf{Hit@k (Normalized): }Whether the normalized version of the true password appears in the top-k predictions.
    \item \textbf{Hit@k (SHA-256): }Whether the exact SHA-256 hash of the true password matches any predicted guess's hash.
    \item \textbf{Coverage: }The proportion of user profiles for which at least one non-empty guess was produced.
\end{itemize}

\subsection{Observations}
\begin{itemize}
    \item \textbf{TinyLlama }achieved the highest Hit@10 (Normalized) among LLMs at 1.34\%, but failed to crack any SHA-256 matches.
    \item \textbf{Flan-T5-Small }maintained consistent but low accuracy (0.57\%) across all Hit@k metrics.
    \item \textbf{Traditional models }such as Rule-Based and Name-Year Fusion achieved substantially higher Hit@k scores, confirming their practical superiority in targeted cracking scenarios.
    \item •	None of the LLMs outperformed the weakest traditional method in terms of exact password matching.
\end{itemize}

These findings suggest that while LLMs can imitate password-like text generation, they lack precision and contextual alignment necessary for high-success-rate password guessing. In contrast, traditional methods, although simpler, exploit deterministic rules and user tendencies more effectively in this context.

\section{Discussion of Results}
The comparative evaluation between Large Language Models (LLMs) and traditional password cracking techniques reveals several critical insights into their respective strengths and limitations.\\

Firstly, the LLMs—namely \textbf{TinyLlama, Falcon-RW-1B} and \textbf{Flan-T5-Small} demonstrated poor performance in actual password matching, with Hit@10 (Normalized) scores remaining below 1.5\% and zero success in exact hash-based matching (Hit@k SHA-256). This highlights a major shortcoming: although LLMs can produce syntactically plausible password-like sequences, they struggle to generate exact matches that align with real user behavior. Despite being fed rich contextual prompts—including name, birthdate, hobbies, and email address—the models failed to learn transformation patterns (e.g., appending years, inserting special characters, or capitalizing certain words) that are often employed in real-world passwords.\\

In contrast, \textbf{traditional methods} such as Rule-Based, Combinator, and Name-Year Fusion exhibited significantly higher Hit@k scores. These techniques rely on domain-specific rules, pre-defined patterns, and user-specific substitutions to guess passwords with greater accuracy. The Rule-Based method, for instance, achieved a Hit@10 accuracy of 33.07\%, vastly outperforming all LLMs. Even simpler heuristics like Mutated Username and Email-Based guessing showed respectable success, proving that deterministic approaches grounded in human password habits remain highly effective.\\

The disparity in performance may be attributed to how LLMs generalize. Models like TinyLlama and Flan-T5 are trained on natural language text, not on structured password generation patterns. Consequently, they often generate overgeneralized guesses that do not match the idiosyncrasies of human password creation. Moreover, the absence of fine-tuning on real password datasets or curated wordlists like rockyou.txt likely further limited their ability to internalize password structure and entropy reduction strategies.\\

Additionally, while LLMs offer scalability and automation, their current performance suggests that they are not yet viable replacements for traditional cracking in digital forensics or penetration testing. However, with targeted fine-tuning and prompt engineering, there may be potential for improvement.\\

Overall, the results underscore a key finding: \textbf{current LLMs, when used out-of-the-box}, lack the specificity and structural modeling required for effective password guessing—especially when compared to even basic traditional methods. This finding challenges the assumption that large-scale language understanding naturally translates into predictive accuracy in specialized security tasks like password cracking.\\

\section{Conclusion and Future Work}
This study presented a systematic evaluation of the password guessing capabilities of Large Language Models (LLMs) compared to traditional cracking techniques. Using a synthetically generated dataset of 20,000 user profiles with contextual attributes such as names, usernames, birthdates, and hobbies, we assessed the effectiveness of three open-access LLMs—TinyLlama, Falcon-RW-1B, and Flan-T5-Small—on their ability to generate accurate password guesses.\\

Our findings reveal that despite the growing popularity of LLMs in natural language processing tasks, their performance in password cracking remains significantly inferior to traditional rule-based and heuristic methods. The best-performing LLM, Flan-T5-Small, achieved only 0.57\% accuracy at Hit@10 (Normalized), while none of the LLMs succeeded in matching even a single password using hash-based verification. In contrast, traditional methods such as Rule-Based and Combinator approaches consistently demonstrated higher success rates, with Hit@10 accuracies exceeding 30\% in some cases.\\

These results underscore an important conclusion: \textbf{LLMs, in their current state and without domain-specific fine-tuning, are not suitable substitutes for traditional password cracking tools}. Their tendency to generate plausible yet imprecise guesses, and their failure to model real-world password transformation patterns, limit their practical utility in cybersecurity applications like digital forensics or penetration testing.\\

Future work can explore several avenues to bridge this gap. First, fine-tuning LLMs on large-scale password datasets—while respecting ethical and legal constraints—may significantly improve performance. Second, integrating user-specific transformation patterns (e.g., appending birth years or favorite digits) through prompt engineering or few-shot learning could make the outputs more targeted. Additionally, hybrid approaches that combine the generative flexibility of LLMs with rule-based filtering mechanisms may offer a balanced compromise between creativity and precision.\\

As password security continues to be a critical frontier in cybersecurity, understanding the limitations and opportunities of AI-driven techniques is essential. This study contributes to that understanding by empirically showing where LLMs fail—and how traditional intelligence still prevails in password cracking.

\bibliographystyle{IEEEtran}
\bibliography{references}

\end{document}